\documentclass{llncs}

\usepackage{mathptmx}       
\usepackage{helvet}         
\usepackage{courier}        
\usepackage{type1cm}        
\usepackage{makeidx}         
\usepackage{graphicx}        
\usepackage{multicol}        
\usepackage[bottom]{footmisc}
\usepackage[boxed]{algorithm}
\usepackage{algorithmic}

\newcommand{\eat}[1]{}

\newcommand{\ie}{{\em i.e.}}
\newcommand{\eg}{{\em e.g.}}

\newcommand{\err}{\varepsilon}

\begin{document}

\title{Data Fusion: Resolving Conflicts from Multiple Sources}
\author{Xin Luna Dong\inst{1}
 \and Laure Berti-Equille\inst{2} \and Divesh Srivastava\inst{3}}
\authorrunning{Xin Luna Dong et al.}
\institute{Google Inc., \email{lunadong@google.com}
\and Institut de Recherche pour le Developpement (IRD),  
\email{Laure.Berti@ird.fr} \\
\and AT\&T Labs-Research, \email{divesh@research.att.com}}
%
%
\maketitle

\abstract{\small Many data management applications, such as setting up Web portals, managing enterprise data, managing community data, and sharing scientific data, require integrating data from multiple sources. Each of these sources provides a set of values and different sources can often provide conflicting values. To present quality data to users, it is critical to resolve conflicts and discover values that reflect the real world; this task is called {\em data fusion}. This paper describes a novel approach that 
finds true values from conflicting information when there are a large number of sources, among which some may copy from others. We present a case study on real-world data showing that the described algorithm can significantly improve accuracy of truth discovery and is scalable when there are a large number of data sources.
}

\section{Introduction}
\label{sec:intro}
The amount of useful information available on the Web has been 
growing at a dramatic pace in recent years.
In a variety of domains, such as science, business, technology, arts,
entertainment, politics, government, sports, tourism, there are a
huge number of data sources that seek to provide information 
to a wide spectrum of information users.
In addition to enabling the availability of useful information,
the Web has also eased the ability to publish and spread false
information across multiple sources.
Widespread availability of conflicting information (some true,
some false) makes it hard to separate the wheat from the chaff.
Simply using the information that is asserted by the largest number
of data sources (\ie, naive voting) is clearly inadequate since biased (and even 
malicious) sources abound, and plagiarism (\ie, copying without
proper attribution) between sources may be widespread.
{\em Data fusion} aims at resolving conflicts from different
sources and find values that reflect the real world. 

Ideally, when applying voting, we would like to give a higher vote to
more trustworthy sources and ignore copied information;
however, this raises many challenges. 
First, we often do not know {\em a priori} the trustworthiness
of a source and that depends on how much of its provided
data are correct, but the correctness of data, on the
other hand, needs to be decided 
by considering the number and trustworthiness of the providers;
thus, it is a chicken-and-egg problem. Second, in many applications 
we do not know how each source obtains its data,
so we have to discover copiers from a snapshot of data.
The discovery is non-trivial: sharing common data does not in itself
imply copying--accurate sources can also share a lot of
independently provided correct data; 
not sharing a lot of common data does not in itself
imply no-copying--a copier may copy only a small fraction of
data from the original source; even when we decide that two
sources are dependent, it is not always obvious which one is a copier.
Third, a copier can also provide some data 
by itself or verify the correctness of some of the copied data, 
so it is inappropriate to ignore all data it provides. 

In this paper, we present novel approaches for data fusion. 
First, we consider {\em copying} between 
data sources in truth discovery. Our technique considers 
not only whether two sources share the same
values, but also whether the shared values are true or false. 
Intuitively, for a particular object, there are often multiple distinct
false values but usually only one true value. Sharing the same true value
does not necessarily imply copying between sources; however, sharing
the same false value is typically a low-probability event when the sources 
are fully independent. Thus, if two data sources share a lot of
false values, copying is more likely. Based on this analysis,
we describe Bayesian models that compute the probability of 
copying between pairs of data sources and take the result
into consideration in truth discovery.

Second, we also consider {\em accuracy} in voting: 
we trust an accurate data source more and give values 
that it provides a higher weight. This method requires identifying
not only if two sources are dependent, but also which source is
the copier. Indeed, accuracy in itself is a clue of 
direction of copying:
given two data sources, if the accuracy of their common data 
is highly different from that of one of the sources,
that source is more likely to be a copier.

\begin{table}[t]
\caption{\label{tbl:motivating} The motivating example: five data 
sources provide information on the affiliations of five researchers.
Only $S_1$ provides all true values.}
\begin{center}
\begin{tabular}{|c|c|c|c|c|c|}
\hline
& $S_1$ & $S_2$ & $S_3$ & $S_4$ & $S_5$ \\
\hline
$Stonebraker$ & MIT & Berkeley & MIT & MIT & MS \\
$Dewitt$ & MSR & MSR & UWisc & UWisc & UWisc \\
$Bernstein$ & MSR & MSR & MSR & MSR & MSR \\
$Carey$ & UCI & AT\&T & BEA & BEA & BEA \\
$Halevy$ & Google & Google & UW & UW & UW \\
\hline
\end{tabular}
\end{center}
\vspace{-.4in}
\end{table}
\begin{example}
\label{eg:motivating}
Consider the five data sources in 
Table~\ref{tbl:motivating}. They provide information
on affiliations of five researchers and only
$S_1$ provides all correct data.
Sources $S_4$ and $S_5$ copy their data from $S_3$,
and $S_5$ introduces certain errors during copying.

First consider the three sources $S_1, S_2,$ and $S_3$.
For all researchers except {\em Carey}, a naive voting
on data provided by these three sources can find the correct affiliations.
For {\em Carey}, these sources provide three different affiliations,
resulting in a tie. However, if we take into account
that the data provided by $S_1$ is more accurate (among the
rest of the 4 researchers, $S_1$ provides all correct affiliations,
whereas $S_2$ provides 3 and $S_3$ provides only 2 correct affiliations), 
we will consider {\em UCI} as most likely to be the correct value.

Now consider in addition sources $S_4$ and $S_5$. 
Since the affiliations provided by $S_3$ are copied by
$S_4$ and $S_5$, naive voting would consider them as the majority
and so make wrong decisions for three researchers. 
Only if we ignore the values provided by $S_4$ and $S_5$,
we will be able to again decide the correct affiliations.
Note however that identifying the copying relationships is
not easy: while $S_3$ shares 5 values with $S_4$ and
4 values with $S_5$, $S_1$ and $S_2$ also share 3 values,
more than half of all values.  
If we knew which values are true and which are false, 
we would suspect copying between $S_3$, $S_4$ and $S_5$,
because they provide the same false values.
On the other hand, we would suspect the copying between 
$S_1$ and $S_2$ much less, as they
share only true values.
\end{example}

The structure of the rest of the paper is as follows. 
Section~\ref{sec:accu} presents how we can leverage 
source accuracy in data fusion. Section~\ref{sec:depen}
presents how we can leverage copying relationships in data fusion.
Section~\ref{sec:case} presents a case study of
these techniques on a real-world data set,
and Section~\ref{sec:summary} concludes.

\section{Fusing Sources Considering Accuracy}
\label{sec:accu}
We first formally describe the data fusion problem and
describe how we leverage the trustworthiness of sources
in truth discovery. In this section we assume no-copying
between data sources and defer discussion on copying 
to the next section.

\subsection{Data Fusion}
We consider a set of {\em data sources} ${\cal S}$ and
a set of {\em objects} ${\cal O}$. An object represents
a particular aspect of a real-world entity, such as 
the affiliation of a researcher; in a relational database, an object
corresponds to a cell in a table. For each object $O \in {\cal O}$,
a source $S \in {\cal S}$ can (but not necessarily) 
provide a {\em value}. Among different values
provided for an object, one correctly describes the real world and is
{\em true}, and the rest are {\em false}.
\eat{Table~\ref{tbl:notation} lists the variables
and parameters we use in this paper and we shall explain each of them
at the time of use.
}In this paper we solve the following problem: given a snapshot of 
data sources in $\cal S$, decide the true value 
for each object $O \in \cal O$.

We note that a value provided by a data source
can either be atomic, or a set or list of atomic values 
(\eg, author list of a book). 
In the latter case, we consider the value as true if the atomic values
are correct and the set or list is complete 
(and order preserved for a list). This
setting already fits many real-world applications and 
we refer our readers to~\cite{ZRHG12} for solutions that 
treat a set or list of values as multiple values.

We consider a core case that satisfies 
the following two conditions (relaxation of these assumptions is discussed
in~\cite{DBS09a}): 
\begin{itemize}
  \item {\em Uniform false-value distribution:} For each object, 
    there are multiple false values in the underlying domain and 
    an independent source has the same probability of providing each of them.
  \item {\em Categorical value:} For each object, values that do not match
    exactly are considered as completely different. 
\end{itemize}

Note that this problem definition focuses on {\em static}
information that does not evolve over time, such as 
authors and publishers of books, and
we refer our readers to~\cite{DBS09b} for data fusion for
evolving values.

\subsection{Accuracy of a Source} 
Let $S \in \cal S$ be a data source. The {\em accuracy} of $S$,
denoted by $A(S)$, is the fraction of true values provided by $S$;
it can also be considered as the probability that a value
provided by $S$ is the true value. 

Ideally we should compute the accuracy of a source as it is defined;
however, in real applications we often do not know for sure
which values are true, especially among values that are provided
by similar number of sources. Thus, we compute the accuracy 
of a source as the average probability of its values being true
(we describe how we compute such probabilities shortly).
Formally, let $\bar V(S)$ be the values provided by $S$
and denote by $|\bar V(S)|$ the size of $\bar V(S)$. 
For each $v \in \bar V(S)$, we denote by $P(v)$ the probability
that $v$ is true. We compute $A(S)$ as follows.

\begin{equation}
\label{equ:accuracy}
A(S) = {\Sigma_{v \in \bar V(S)} P(v) \over |\bar V(S)|}.
\end{equation}

We distinguish {\em good} sources from {\em bad} ones:
a data source is considered to be good if for each object
it is more likely to provide the true value than any 
{\em particular} false value; otherwise, it is considered to be bad.
Assume for each object in $\cal O$ the number of
false values in the domain is $n$. Then, in the core case, 
the probability that $S$ provides a true value
is $A(S)$ and that it provides a particular false value is
$1-A(S) \over n$. So $S$ is good if $A(S)>{1-A(S) \over n}$
(\ie, $A(S) > {1 \over 1+n}$).
We focus on good sources in the rest of this paper,
unless otherwise specified.

\subsection{Probability of a Value Being True}
Now we need a way to compute the probability that a value
is true. Intuitively, the computation should consider both how many sources
provide the value and accuracy of those sources. 
We apply a Bayesian analysis for this purpose.

Consider an object $O \in \mathcal{O}$.
Let $\mathcal{V}(O)$ be the domain of $O$, including one
true value and $n$ false values. \eat{; that is, the set of
all possible values on $O$. The size of
$\mathcal{V}(O)$ is $n+1$: $n$ false values and one true value.
}Let $\bar S_o$ be the sources that provide information on $O$.
For each $v \in \mathcal{V}(O)$, we denote by 
$\bar S_o(v) \subseteq \bar S_o$ the set of
sources that vote for $v$ ($\bar S_o(v)$ can be empty). 
We denote by $\Psi(O)$ the observation of which value each
$S \in \bar S_o$ votes for $O$. 

To compute $P(v)$ for $v \in \mathcal{V}(O)$,
we need to first compute the probability of $\Psi(O)$ conditioned on
$v$ being true. 
This probability should be that of sources in 
$\bar S_o(v)$ each providing the true value and 
other sources each providing a particular false value:

\begin{eqnarray}
\nonumber
Pr(\Psi(O) | v\ \mbox{true}) 
&=& \Pi_{S \in \bar S_o(v)}A(S)\cdot\Pi_{S \in \bar S_o\setminus\bar S_o(v)}{1-A(S) \over n} \\
&=& \Pi_{S \in \bar S_o(v)}{nA(S) \over 1-A(S)}\cdot\Pi_{S \in \bar S_o}{1-A(S) \over n}.
\label{eqn:disPr}
\end{eqnarray}

Among the values in $\mathcal{V}(O)$, there is one and only one true value.
Assume our {\em a priori} belief of each value being true is the same, denoted by 
$\beta$. We then have

\begin{equation}
Pr(\Psi(O)) 
= \sum_{v \in \mathcal{V}(O)}\left(\beta\cdot\Pi_{S \in \bar S_o(v)}{nA(S) \over 1-A(S)} \cdot \Pi_{S \in \bar S_o}{1-A(S) \over n}\right).
\end{equation}

Applying the Bayes Rule leads us to

\begin{equation}
P(v) = Pr(v\ \mbox{true} | \Psi(O))  
= {\Pi_{S \in \bar S_o(v)}{nA(S) \over 1-A(S)} \over 
\sum_{v_0 \in \mathcal{V}(O)} \Pi_{S \in \bar S_o(v_0)} {nA(S) \over 1-A(S)}
}.
\label{equ:valuePr}
\end{equation}

To simplify the computation,
we define the {\em confidence} of $v$, denoted by $C(v)$, 
as $C(v)= \sum_{S \in \bar S_o(v)}\log{nA(S) \over 1-A(S)}$.
If we define the {\em accuracy score} of a data source $S$ as
$A'(S) = \log{nA(S) \over 1-A(S)}$, we have 
$C(v)=\sum_{S \in \bar S_o(v)}A'(S)$.
So we can compute
the confidence of a value by summing up the accuracy scores of its
providers. Finally, we can compute the probability of each value as 
$P(v)={2^{C(v)} \over \sum_{v_0 \in {\cal V}(O)}2^{C(v_0)}}$.
A value with a higher confidence has a higher probability to be true;
thus, rather than comparing vote counts, we can just compare confidence
of values. The following theorem shows three nice properties 
of Equation~(\ref{equ:valuePr}).
\begin{theorem}
Equation~(\ref{equ:valuePr}) has the following properties:
\begin{enumerate}
  \item If all data sources are good and have the same accuracy,
	when the size of $\bar S_o(v)$ increases, $C(v)$ increases;
  \item Fixing all sources in $\bar S_o(v)$ except $S$, when $A(S)$ increases for $S$,
	$C(v)$ increases. 
  \item If there exists $S \in \bar S_o(v)$ such that $A(S)=1$ and no $S' \in \bar S_o(v)$
        such that $A(S')=0$,
	$C(v)=+\infty$; if there exists $S \in \bar S_o(v)$ such that $A(S)=0$ and no
	$S' \in \bar S_o(v)$ such that $A(S')=1$,
	$C(v)=-\infty$. 
\end{enumerate}
\end{theorem}

Note that the first property is actually a justification for the
naive voting strategy when all sources have the same accuracy. 
The third property shows that we should
be careful not to assign very high or very low accuracy to a data source,
which has been avoided by defining the accuracy of a source 
as the average probability of its provided values.

\begin{example}
Consider $S_1, S_2$ and $S_3$ in Table~\ref{tbl:motivating}
and assume their accuracies are .97, .6, .4 respectively.
Assuming there are 5 false values in the domain (\ie, $n=5$),
we can compute the accuracy score of each source as follows.
For $S_1$, $A'(S_1)=\log{5*.97 \over 1-.97}=4.7$;
for $S_2$, $A'(S_2)=\log{5*.6 \over 1-.6}=2$;
and for $S_3$, $A'(S_3)=\log{5*.4 \over 1-.4}=1.5$.

Now consider the three values provided for {\em Carey}.
Value {\em UCI} thus has confidence 8, {\em AT\&T} has 
confidence 5, and {\em BEA} has confidence 4.
Among them, {\em UCI} has the highest confidence and 
so the highest probability to be true. Indeed, its probability 
is ${2^8 \over 2^8+2^5+2^4+(5-2)*2^0}=.9$. 
\end{example}

Computing value confidence requires knowing accuracy 
of data sources, whereas computing source accuracy 
requires knowing value probability. There is an
inter-dependence between them and
we solve the problem by computing them iteratively.
We give details of the iterative algorithm in Section~\ref{sec:depen}.

\section{Fusing Sources Considering Copying}
\label{sec:depen}
Next, we describe how we detect copiers and leverage the
discovered copying relationships in data fusion.

\subsection{Copy Detection}
We say that there exists {\em copying} between two data sources 
$S_1$ and $S_2$ if they derive the same part of their data 
directly or transitively from a common source 
(can be one of $S_1$ and $S_2$). 
Accordingly, there are two types of data sources:
{\em independent sources} and {\em copiers}. 
An {\em independent source} provides all values independently. 
It may provide some erroneous values because of incorrect knowledge of the
real world, mis-spellings, etc.
A {\em copier} copies a part (or all) of its data from other 
sources (independent sources or copiers). It can copy from multiple sources
by union, intersection, etc., and as we focus on a snapshot of
data, cyclic copying on a particular object is impossible. 
In addition, a copier may
revise some of the copied values or add additional values; 
though, such revised and added values are considered
as independent contributions of the copier.

To make our models tractable, we consider only {\em direct} copying.
In addition, we make the following assumptions.
\begin{itemize}
  \item {\em Assumption 1 (Independent values).} The values that are independently provided 
    by a data source on different objects are independent of each other.
  \item {\em Assumption 2 (Independent copying).} 
    The copying between a pair of data sources is independent of the 
    copying between any other pair of data sources.
  \item {\em Assumption 3 (No mutual copying).} There is no mutual 
    copying between a pair of sources; that is, $S_1$ copying from
    $S_2$ and $S_2$ copying from $S_1$ do not happen at the same time.
\end{itemize}

Our experiments on real world data 
show that the basic model already obtains high accuracy
and we refer our readers to~\cite{DBH+10a} for how we can
relax the assumptions. 
We next describe the basic copy-detection model.

Consider two sources $S_1, S_2 \in \cal S$.
We apply Bayesian analysis to compute the probability of copying between
$S_1$ and $S_2$ given observation of their data.
For this purpose, we need to compute the probability of the observed data,
conditioned on independence of or copying between the sources.
We denote by $c\ (0 < c \leq 1)$ the probability that a value provided by a copier is copied.
We bootstrap our algorithm by setting $c$
to a default value initially and iteratively 
refine it according to copy detection results.

In our observation, we are interested in three sets of objects:
$\bar O_t$, denoting the set of objects on which $S_1$ and $S_2$
provide the same true value,
$\bar O_f$, denoting the set of objects on which they provide the same false value,
and $\bar O_d$, denoting the set of objects on which they provide
different values ($\bar O_t \cup \bar O_f \cup \bar O_d \subseteq \mathcal{O}$).
Intuitively, two independent sources providing the same false value is 
a low-probability event; thus, if we fix 
$\bar O_t \cup \bar O_f$ and $\bar O_d$, the more common false values 
that $S_1$ and $S_2$ provide, the more likely that they are dependent. 
On the other hand, if we fix 
$\bar O_t$ and $\bar O_f$, the fewer objects on which $S_1$ and $S_2$ provide
different values, the more likely that they are dependent.
We denote by $\Phi$ the observation of $\bar O_t, \bar O_f, \bar O_d$
and by $k_t, k_f$ and $k_d$ their sizes respectively.
We next describe how we compute the conditional probability of 
$\Phi$ based on these intuitions.

We first consider the case where 
$S_1$ and $S_2$ are independent, denoted by $S_1 \bot S_2$.
Since there is a single true value, the
probability that $S_1$ and $S_2$ provide the same true value for object $O$ is

\begin{equation}
Pr(O \in \bar O_t | S_1 \bot S_2) = A(S_1)\cdot A(S_2).
\end{equation}

On the other hand, the
probability that $S_1$ and $S_2$ provide the same false value for $O$ is

\begin{equation}
Pr(O \in \bar O_f | S_1 \bot S_2) = n\cdot {1-A(S_1) \over n}\cdot{1-A(S_2) \over n} = {(1-A(S_1))(1-A(S_2)) \over n}.
\end{equation}

Then, the probability that $S_1$ and $S_2$ provide
different values on an object $O$, denoted by $P_d$ for convenience, is

\begin{equation}
\label{eqn:d1}
Pr(O \in \bar O_d | S_1 \bot S_2) = 1-A(S_1)A(S_2)-{(1-A(S_1))(1-A(S_2)) \over n} = P_d.
\end{equation}

Following the {\em Independent-values} assumption, the conditional probability of
observing $\Phi$ is

\begin{equation}
Pr(\Phi | S_1 \bot S_2) = 
{A(S_1)^{k_t}A(S_2)^{k_t}(1-A(S_1))^{k_f}(1-A(S_2))^{k_f}P_d^{k_d} \over n^{k_f}}.
\end{equation}

We next consider the case when $S_2$ copies from $S_1$,
denoted by $S_2 \to S_1$.
There are two cases where $S_1$ and $S_2$ provide the same value $v$
for an object $O$. First, 
with probability $c$, $S_2$ copies $v$ from $S_1$ and so 
$v$ is true with probability $A(S_1)$ and false with
probability $1-A(S_1)$. Second, with probability $1-c$,
the two sources provide $v$ independently and so its probability of being
true or false is the same as in the case where $S_1$ and $S_2$ are independent.
Thus, we have

\begin{eqnarray}
Pr(O \in \bar O_t | S_2 \to S_1) &=& A(S_1) \cdot c + A(S_1)\cdot A(S_2)\cdot (1-c), \\
Pr(O \in \bar O_f | S_2 \to S_1) &=& (1-A(S_1))\cdot c + {(1-A(S_1))(1-A(S_2)) \over n} \cdot (1-c). 
\end{eqnarray}

Finally, the probability that $S_1$ and $S_2$ provide different
values on an object is that of $S_1$ providing a value independently
and the value differs from that provided by $S_2$:

\begin{equation}
\label{eqn:d2}
Pr(O \in \bar O_d | S_2 \to S_1) = P_d \cdot (1-c).
\end{equation}

We compute $Pr(\Phi | S_2 \to S_1)$ accordingly;
similarly we can also compute $Pr(\Phi | S_1 \to S_2)$.
Now we can compute the probability of $S_1 \bot S_2$
by applying the Bayes Rule. 

\begin{eqnarray}
\nonumber
& & Pr(S_1 \bot S_2 | \Phi) \\
&=& \frac{\alpha Pr(\Phi | S_1 \bot S_2)}
         {\alpha Pr(\Phi | S_1 \bot S_2) +
	  {1-\alpha \over 2}Pr(\Phi | S_1 \to S_2) +
  	  {1-\alpha \over 2}Pr(\Phi | S_2 \to S_1)}.
\label{eqn:bene}
\end{eqnarray}
Here $\alpha=Pr(S_1 \bot S_2) (0<\alpha<1)$ 
is the {\em a priori} probability that two data sources are independent. 
As we have no {\em a priori} preference for copy direction,
we set the {\em a priori} probability for copying in each direction
as $1-\alpha \over 2$.

Equation~(\ref{eqn:bene}) has several nice properties that 
conform to the intuitions we discussed earlier in this section,
formalized as follows.
\begin{theorem}
\label{thm:property}
Let ${\cal S}$ be a set of good independent sources and copiers.
Equation~(\ref{eqn:bene}) has the following three properties on ${\cal S}$.
\begin{enumerate}
  \item Fixing $k_t+k_f$ and $k_d$, when $k_f$ increases, the probability of copying (i.e., $Pr(S_1 \to S_2|\Phi)+Pr(S_2 \to S_1|\Phi)$)
    increases;
  \item Fixing $k_t+k_f+k_d$, when $k_t+k_f$ increases and none of $k_t$ and
    $k_f$ decreases, the probability of copying increases;
  \item Fixing $k_t$ and $k_f$, when $k_d$ decreases, the probability of copying
    increases. 
\end{enumerate}
\end{theorem}

\begin{example}\label{eg:nocopy}
Continue with Ex.\ref{eg:motivating} and consider the possible copying
relationship between $S_1$ and $S_2$. We observe that they
share no false values (all values they share are correct),
so copying is unlikely. With $\alpha=.5, c=.2, A(S_1)=.97, A(S_2)=.6$,
the Bayesian analysis goes as follows.

We start with computation of $Pr(\Phi|S_1 \bot S_2)$.
We have $Pr(O \in \bar O_t | S_1 \bot S_2) = .97*.6=.582$.
There is no object in $\bar O_f$ and we denote by $P_d$ the
probability $Pr(O \in \bar O_f | S_1 \bot S_2)$.
Thus, $Pr(\Phi|S_1 \bot S_2)=.582^3*P_d^2=.2P_d^2$.

Next consider $Pr(\Phi|S_1 \to S_2)$. 
We have $Pr(O \in \bar O_t | S_1 \bot S_2) = .8*.6+.2*.582=.6$ 
and $Pr(O \in \bar O_f | S_1 \to S_2) = .2P_d$.
Thus, $Pr(\Phi|S_1 \to S_2)=.6^3*(.2P_d)^2=.008P_d^2$.
Similarly, $Pr(\Phi|S_2 \to S_1)=.028P_d^2$.

According to Equation~(\ref{eqn:bene}), 
$Pr(S_1 \bot S_2|\Phi)={.5*.2P_d^2 \over .5*.2P_d^2+.25*.008P_d^2+.25*.028P_d^2}=.92$, so independence is very likely.
\end{example}

\subsection{Independent Vote Count of a Value}
Since even a copier can provide some of the values independently, 
we compute the {\em independent} vote for each particular value.
In this process we consider the data sources one by one in some order. 
For each source $S$, we denote by
$\overline{Pre}(S)$ the set of sources that have already been considered
and by $\overline{Post}(S)$ the set of sources that have not been considered
yet. We compute the probability that the value provided by $S$ 
is independent of any source in $\overline{Pre}(S)$ and take it as
the vote count of $S$. The vote count computed in this way
is not precise because if $S$
depends only on sources in $\overline{Post}(S)$ but some of those sources 
depend on sources in $\overline{Pre}(S)$, our estimation still (incorrectly)
counts $S$'s vote. To minimize such
error, we wish that the probability that $S$ depends on 
a source $S' \in \overline{Post}(S)$ and $S'$ depends on a source
$S'' \in \overline{Pre}(S)$ be the lowest. Thus, we use a greedy algorithm
and consider data sources in the following order.

\begin{enumerate}
  \item If the probability of $S_1 \to S_2$ is much higher
than that of $S_2 \to S_1$, we consider $S_1$
as a copier of $S_2$ with probability 
$Pr(S_1 \to S_2|\Phi)+Pr(S_2 \to S_1|\Phi)$ (recall that we assume
there is no mutual-copying) and order $S_2$ before $S_1$. Otherwise, we
consider both directions as equally possible and 
there is no particular order between $S_1$ and $S_2$;
we consider such copying {\em undirectional}.
  \item For each subset of sources between which there is
no particular ordering yet, we sort them
as follows: in the first round, we select a data source 
that is associated with the undirectional copying of the highest
probability ($Pr(S_1 \to S_2|\Phi)+Pr(S_2 \to S_1|\Phi)$); 
in later rounds, each time we select a data source that has
the copying with the maximum probability with 
one of the previously selected sources.
\end{enumerate}

We now consider how to compute the vote count of $v$
once we have decided an order of the data sources.
Let $S$ be a data source that votes for $v$.
The probability that $S$ provides $v$ independently of a source
$S_0 \in \overline{Pre}(S)$ is 
$1-c(Pr(S_1 \to S_0|\Phi)+Pr(S_0 \to S_1|\Phi))$ 
and the probability that $S$ provides $v$ independently of any data source
in $\overline{Pre}(S)$, denoted by $I(S)$, is

\begin{equation}
\label{eqn:vote-bene}
I(S)=\Pi_{S_0 \in \overline{Pre}(S)}(1-c(Pr(S_1 \to S_0|\Phi)+Pr(S_0 \to S_1|\Phi))).
\end{equation}
The total vote count of $v$ is $\sum_{S \in \bar S_o(v)}I(S)$.

Finally, when we consider the accuracy of sources, we compute 
the confidence of $v$ as follows.

\begin{equation}
C(v)=\sum_{S \in \bar S_o(v)}A'(S)I(S).
\end{equation}

In the equation, $I(S)$ is computed by Equation~(\ref{eqn:vote-bene}). 
In other words, we take only the ``independent fraction'' of the
original vote count (decided by source accuracy) from each source.

\subsection{Iterative Algorithm}
We need to compute three measures: accuracy of 
sources, copying between sources, and confidence of
values. Accuracy of a source depends on confidence of values;
copying between sources depends on accuracy of sources and
the true values selected according to the confidence of values; and 
confidence of values depends on both accuracy of and copying 
between data sources.

We conduct analysis of both accuracy and copying
in each round. Specifically, Algorithm {\sc AccuCopy} 
starts by setting the same accuracy for each source
and the same probability for each value, then 
iteratively (1) computes copying based on the 
confidence of values computed in the previous round, (2)
updates confidence of values accordingly,
and (3) updates accuracy of sources accordingly,
and stops when the accuracy of the sources becomes stable.
Note that it is crucial to consider copying between sources
from the beginning;
otherwise, a data source that has been duplicated many times
can dominate the voting results in the first round and make it hard to detect the
copying between it and its copiers (as they share only ``true''
values). Our initial decision on copying is similar to
Equation~(\ref{eqn:bene}) except considering both the possibility
of a value being true and that of the value being false 
and we skip details here.

We can prove that if we ignore source accuracy 
(\ie, assuming all sources have the same
accuracy) and there are a finite number of objects in $\cal O$,
Algorithm~{\sc AccuCopy} cannot change the decision for an object $O$ 
back and forth between two different values forever;
thus, the algorithm converges. 
\begin{theorem}
Let $\cal S$ be a set
of good independent sources and copiers
that provide information on objects in $\cal O$.
Let $l$ be the number of objects in $\cal O$ and
$n_0$ be the maximum number of values provided for an object by $\cal S$.
The {\sc AccuVote} algorithm converges in at most $2l n_0$
rounds on $\cal S$ and $\cal O$ if it ignores source accuracy. 
\end{theorem}

Once we consider accuracy of
sources, {\sc AccuCopy} may not converge: when we select
different values as the true values, the direction of the
copying between two sources can change and in turn suggest
different true values. We stop the process
after we detect oscillation of decided true values.
Finally, we note that the complexity of each round is 
$O(|{\cal O}||{\cal S}|^2\log |{\cal S}|)$.

\section{A Case Study}
\label{sec:case}
We now describe a case study on a real-world data set\footnote{\small 
http://lunadong.com/fusionDataSets.htm.}
extracted by searching computer-science books on 
{\em AbeBooks.com}. For each book, {\em AbeBooks.com} returns
information provided by a set of online bookstores. 
Our goal is to find the list of authors for each book. In the data set
there are 877 bookstores, 1263 books,
and 24364 listings (each listing contains a list of authors on a book
provided by a bookstore). 

We did a normalization of author names and 
generated a normalized form that preserves the order of the
authors and the first name and last name (ignoring the middle name) 
of each author. 
On average, each book has 19 listings; the number of different
author lists after cleaning varies from 1 to 23 and is 4 on average.

\begin{table}[t]
\vspace{-.2in}
\caption	{\label{tbl:type}Different types of errors by
naive voting.}
\vspace{-.4in}
\begin{center}
\begin{tabular}{|c|c|c|c|c|}
\hline
Missing authors & Additional authors & Mis-ordering & Mis-spelling & Incomplete names\\
\hline
23 & 4 & 3 & 2 & 2 \\
\hline
\end{tabular}
\end{center}
\vspace{-.4in}
\end{table}
We used a golden standard that contains 100 randomly selected books and
the list of authors found on the cover of each book.
We compared the fusion results with the golden standard,
considering missing or
additional authors, mis-ordering, misspelling, and missing 
first name or last name as errors; however, we do not report
missing or misspelled middle names. Table~\ref{tbl:type} shows
the number of errors of different types on the selected books
if we apply a naive voting (note that the result
author lists on some books may contain multiple types of errors).

We define {\em precision} of the results as the fraction of objects on which 
we select the true values (as the number of true values we return 
and the real number of true values are both the same as the number of
objects, the {\em recall} of the results is the same as the precision). 
Note that this definition is different from that of accuracy of sources.

\medskip
\noindent
{\bf Precision and Efficiency}
\begin{table}[t]
\caption	{\label{tbl:book}Results on the book data set. For each method,
we report the precision of the results, the run time, and the number of rounds
for convergence. {\sc AccuCopy} and {\sc Copy} obtain a high precision.}
\vspace{-.1in}
\begin{center}
\begin{tabular}{|c|c|c|c|}
\hline
Model & Precision & Rounds & Time (sec) \\
\hline
{\sc Vote} & .71 & 1 & .2 \\
{\sc Sim} & .74 & 1 & .2 \\
{\sc Accu} & .79 & 23 & 1.1 \\
{\sc Copy} & .83 & 3 & 28.3 \\
{\sc AccuCopy} & .87 & 22 & 185.8 \\
{\sc AccuCopySim} & .89 & 18 & 197.5 \\
\hline
\end{tabular}
\end{center}
\vspace{-.4in}
\end{table}
We compared the following data fusion models on this data set.
\begin{itemize}
  \item {\sc Vote} conducts naive voting;
  \item {\sc Sim} conducts naive voting but considers similarity between values;
  \item {\sc Accu} considers accuracy of sources as we described in
	Section~\ref{sec:accu}, but assumes all sources are independent;
  \item {\sc Copy} considers copying between sources as we described in
	Section~\ref{sec:depen}, but assumes all sources have
        the same accuracy;
  \item {\sc AccuCopy} applies the {\sc AccuCopy} algorithm described
	in Section~\ref{sec:depen}, considering both source accuracy
        and copying.
  \item {\sc AccuCopySim} applies the {\sc AccuCopy} algorithm 
        and considers in addition similarity between values.
\end{itemize}

When applicable, we set $\alpha=.2, c=.8, \err=.2$ and $n=100$.
Though, we observed that ranging $\alpha$ from .05 to .5, 
ranging $c$ from .5 to .95, and ranging $\err$ from .05 to .3 
did not change the results much.
We compared similarity of two author lists using 2-gram 
Jaccard distance.

Table~\ref{tbl:book} lists the precision of results of each algorithm.
{\sc AccuCopySim} obtained the best results and improved
over {\sc Vote} by 25.4\%.
{\sc Sim, Accu} and {\sc Copy} each extends {\sc Vote} on a different
aspect; while each of them increased the precision, {\sc Copy} increased it the most. 

To further understand how considering copying and accuracy of sources
can affect our results, 
we looked at the books on which {\sc AccuCopy} and {\sc Vote}
generated different results and manually found the correct authors.
There are 143 such books, 
among which {\sc AccuCopy} gave correct authors for 119 books,
{\sc Vote} gave correct authors for 15 books, and
both gave incorrect authors for 9 books.

Finally, {\sc Copy} was quite efficient 
and finished in 28.3 seconds. It took {\sc AccuCopy} and {\sc AccuCopySim} longer time
to converge (3.1, 3.3 minutes respectively); however, truth discovery
is often a one-time process and so taking a few minutes is reasonable. 

\medskip
\noindent
{\bf Copying and source accuracy:}
\begin{table}[t]
\vspace{-.4in}
\caption{\label{tbl:book_depen}Bookstores that are likely to
be copied by more than 10 other bookstores. For each bookstore we
show the number of books it lists and its accuracy computed by {\sc AccuCopySim}.}
\vspace{-.1in}
\begin{center}
\begin{tabular}{|c|c|c|c|}
\hline
Bookstore & \#Copiers & \#Books & Accuracy \\
\hline
Caiman & 17.5 & 1024 & .55 \\
MildredsBooks & 14.5 & 123 & .88 \\
COBU GmbH \& Co. KG & 13.5 & 131 & .91 \\
THESAINTBOOKSTORE & 13.5 & 321 & .84 \\
Limelight Bookshop & 12 & 921 & .54 \\
Revaluation Books & 12 & 1091 & .76 \\
Players Quest & 11.5 & 212 & .82 \\
AshleyJohnson & 11.5 & 77 & .79 \\
Powell's Books & 11 & 547 & .55 \\
AlphaCraze.com & 10.5 & 157 & .85 \\
\hline 
{\em Avg} & 12.8 & 460 & .75 \\
\hline
\end{tabular}
\end{center}
\vspace{-.4in}
\end{table}
Out of the 385,000 pairs of bookstores, 2916 pairs 
provide information on at least the same 10 books and among them
{\sc AccuCopySim} found 508 pairs that are likely
to be dependent. Among each such pair $S_1$ and $S_2$,
if the probability of $S_1$ depending on $S_2$ is over 2/3 of the
probability of $S_1$ and $S_2$ being dependent, 
we consider $S_1$ as a {\em copier} of $S_2$; otherwise, we consider
$S_1$ and $S_2$ each has .5 probability to be a {\em copier}.
Table~\ref{tbl:book_depen} shows the bookstores whose information
is likely to be copied by more than 10 bookstores.
On average each of them provides information on 460 books and has accuracy .75.
Note that among all bookstores, on average each provides information on
28 books, conforming to the intuition
that small bookstores are more likely to copy data from large ones.
Interestingly, when we applied {\sc Vote}
on only the information provided by bookstores in Table~\ref{tbl:book_depen},
we obtained a precision of only .58, showing that bookstores that
are large and copied often actually can make a lot of mistakes. 

\begin{table}[t]
\caption{\label{tbl:book_accu}Difference between accuracy of sources
computed by our algorithms and the sampled accuracy on the golden standard.
The accuracy computed by {\sc AccuCopySim} is the closest 
to the sampled accuracy.}
\vspace{-.1in}
\begin{center}
\begin{tabular}{|c||c||c|c|c|c|}
\hline
 & Sampled & {\sc AccuCopySim} & {\sc AccuCopy} & {\sc Accu} \\
\hline
Average source accuracy & .542 & .607 & .614 & .623 \\
\hline
Average difference & - & .082 & .087 & .096 \\
\hline 
\end{tabular}
\end{center}
\vspace{-.5in}
\end{table}
Finally, we compare the source accuracy computed by our algorithms
with that sampled on the 100 books in the
golden standard. Specifically, there were 46 bookstores that provide
information on more than 10 books in the golden standard.
For each of them we computed the {\em sampled accuracy} as the
fraction of the books on which the bookstore provides the same author
list as the golden standard. Then, for each bookstore
we computed the difference between its accuracy
computed by one of our algorithms and the sampled accuracy
(Table~\ref{tbl:book_accu}). The source accuracy
computed by {\sc AccuCopySim} is the closest to the sampled accuracy, 
indicating the effectiveness of our model on computing source accuracy
and showing that  considering copying between sources
helps obtain better source accuracy.

\section{Related Work and Conclusions}
\label{sec:summary}
This paper presented how to improve truth discovery by 
analyzing accuracy of sources and detecting copying between sources.
We describe Bayesian models that discover copiers by analyzing values 
shared between sources. A case study
shows that the presented algorithms can significantly
improve accuracy of truth discovery and are scalable 
when there are a large number of data sources.

Our work is closely related to {\em Data Provenance}, which has been a topic of research 
for a decade~\cite{BCTV08,DF08}. Whereas research on data provenance is focused on
how to represent and analyze available provenance information, our work on copy detection helps
detect provenance and in particular copying relationships between dependent data sources.

Our work is also related to analysis of trust and authoritativeness of 
sources~\cite{AG10,BRRT05,pagerank,Kleinberg98,KSG03,SL03} by link analysis or source 
behavior in a P2P network. Such trustworthiness is not directly related to source accuracy. 

Finally, various fusion models have been proposed in the literature.
A comparison of them is presented in~\cite{LDL+12} on two real-world Deep Web data sets,
showing advantages of considering source accuracy together with copying in data fusion.

{\small
\bibliographystyle{abbrv}
\bibliography{../base}  
}
\end{document}